# Lagrangian kinematics of steep waves up to the inception of a spilling breaker

Lev Shemer[1] and Dan Liberzon[2]

[1] School of Mech. Eng., Tel Aviv University, Tel Aviv 69978, Israel
[2] Faculty of Civil and Environ. Eng., Technion, Haifa 3200, Israel

Horizontal Lagrangian velocities and accelerations at the surface of steep water-waves are studied by Particle Tracking Velocimetry (PTV) for gradually increasing crest heights up to the inception of a spilling breaker. Localized steep waves are excited using wavemaker-generated Peregrine breather-type wave trains. Actual crest and phase velocities are estimated from video recorded sequences of the instantaneous wave shape as well as from surface elevation measurements by wave gauges. Effects of nonlinearity and spectral width on phase velocity, as well as relation between the phase velocity and crest propagation speed are discussed. The inception of a spilling breaker is associated with the horizontal velocity of water particles at the crest attaining that of the crest, thus confirming the kinematic criterion for inception of breaking.

**INTRODUCTION**

The concept of wave breaking, although intuitively clear, defies exact and generally accepted definition. While breaking of shoaling waves approaching shore is more familiar, sufficiently steep waves also break in water of large and intermediate depth; the breaking of those waves is of cardinal importance for the energy balance in the ocean. Possible definitions of the wave breaking phenomenon were recently reviewed in Babanin[1] and Perlin *et al*.[2] Distinction is made between spilling and plunging breakers; additional types of breakers are also considered. The spilling breakers usually do not generate foam on the water surface; their shape is strongly affected by viscosity and surface tension effects.[3-6] Plunging breakers are characterized by wave overturning, so that the surface elevation profile ceases to be single-valued. In this process air is entrapped and foam appears on water surface. The breaking process is physically characterized by dissipation of a substantial part of the wave energy, mostly to turbulent kinetic energy of water velocity fluctuations and eventually to heat. Nevertheless, up to the inception of a plunging breaker the wave field can usually be described by potential flow models with a reasonable accuracy.

Probably the most important unresolved question in breaking waves' mechanics is determining the conditions required for waves to break. Numerous breaking criteria were suggested over the years; these criteria may be divided in a broad sense into three types: geometric, kinematic and dynamic.[6,1,2] Various geometric criteria are related to the wave shape on the verge of breaking. Stokes showed that the highest possible wave has steepness of *ak* ≈0.443, with surface elevation forming an angle of 120°. For a random sea, the actual values of the wave steepness are substantially lower than the critical Stokes limit; Ochi and Tsai[7] suggested that waves in sea break when $H/gT^2 \geq 0.02$, *H* and *T* being the local wave height and period respectively. Bonmarin[8] and Babanin *et al.*[9,10] invoke the horizontal wave asymmetry, as well as skewness (the vertical asymmetry), as parameters affecting wave breaking. Chalikov and Sheinin[11] assumed in their numerical computations that the wave breaks when the instantaneous slope becomes vertical.

Since for sufficiently long water waves gravity is the only restoring force, Phillips[12] argued that that the maximum possible negative value of the Lagrangian vertical acceleration is $a_v=-g$. Although physically straightforward, this dynamic criterion of Phillips remains unsupported either experimentally or numerically. In fact, as shown by Longuet-Higgins,[13] the maximum negative value of the Lagrangian vertical acceleration in Stokes 120° corner flow is only $a_v=-g/2$. Computations of the vertical component of the Lagrangian acceleration at the crest of a steep wave were performed by Shemer.[14] Analysis accurate to the 3rd order in wave steepness was carried out for deterministic nonlinear focused unidirectional wave groups with wide spectra that were studied experimentally in a 300 m long wave tank by Shemer *et al.*[15] Computations were performed for conditions corresponding to experimental conditions where the single steep wave was either on a verge of breaking, or actual breaking was observed. It was found that accounting for higher order terms significantly increases the 'apparent' vertical acceleration $\partial^2\eta/\partial t^2$ at the free surface $\eta$ as compared to the linear calculation due to increased weight of higher harmonics. Nevertheless, the Lagrangian acceleration accurate to the 3rd order does not differ notably from the linear result since the convective terms that contribute at the 2nd and higher orders are positive at the crest of the steepest wave, effectively canceling the negative higher order contributions to $\partial^2\eta/\partial t^2$. The negative vertical accelerations at the crest thus do not significantly exceed *g*/3.

This failure of the dynamic criterion suggests that a closer look at the kinematic wave breaking criteria that relate water velocities at the surface with those of wave propagation is appropriate. The kinematic condition states that wave breaks when the water particle velocity at the crest of the wave exceeds the crest velocity that is often represented by the phase velocity $c_p$ at the dominant wave frequency. Alternatively, since the envelope of a narrow-banded group propagates with the group velocity $c_g$, the value of $c_g$ sometimes is taken as the characteristic wave velocity. For example, Tulin and Landrini[16] state that as long as the fluid particle velocity at the wave's crest is lower than $c_g$, the wave does not break.

In numerous experiments on breaking waves initially monochromatic waves were generated by a wavemaker; these waves either break fast as a result of initially high steepness,[17] or at a later stage in the process of evolution due to wave instability.[18,19] In such experiments plunging breakers were usually observed. Perlin *et al.*[17] and Chang and Liu[19] report on measured maximum horizontal velocities in the plunging jet exceeding phase velocity $c_p$, while simultaneous measurements of the horizontal velocity $u$ and of the surface elevation by Melville and Rapp[18] seem to indicate that the values of $u$ remain below $c_p$ even when large velocity excursions were observed in the breaking waves.

It is well known that nonlinearity strongly affects the kinematics of water particles under steep waves as well as the crest velocity.[20] Laboratory measurements of kinematics under steep waves were performed by Jensen *et al.*[21] and Grue and Jensen;[22] when possible, results were compared with those accumulated during field experiments. Nonlinear dependence of surface velocity on wave's steepness was suggested on the basis of those studies. The measured surface velocity of breaking waves was reported to be significantly below the phase velocity of the dominant waves.

It should be stressed that for a wider spectrum, neither $c_p$ nor $c_g$ corresponds to the highest wave crest propagation velocity even if nonlinear effects are neglected.[14] Hence the kinematic criterion for wave breaking should be applied to the relation between the water particle and actual crest velocities.[23] This criterion was examined experimentally by Stansell and MacFarlane[24] who applied various direct and indirect methods to determine the effective phase velocity for the conditions prevailing at wave breaking. Measurements performed for both plunging and spilling breakers indicated that the maximum horizontal velocity at the water surface never reaches the local crest speed. Three-dimensional breaking waves were studied by Wu and Nepf.[25] Horizontal velocity exceeding local wave phase velocity was shown to be a good indication for spilling breakers occurrence, while the horizontal velocity is exceeding $1.5c_p$ is the indication of the occurrence of a plunging breaker. Qiao and Duncan[5] performed simultaneous measurements of the crest velocity and of the horizontal velocity near water surface under a gentle spilling breaker. Although their results exhibit considerable scatter, they seem to indicate that that the maximum horizontal velocity of fluid particle at the surface of a spilling breaker may exceed that of the crest.

The present study deals with the kinematics of breaking waves as well as of waves on verge of breaking and differs from the previous experimental investigations of in several important aspects. The first difference is associated with generation of breaking waves in a laboratory tank. Two approaches to generate breaking waves in a laboratory tank were employed in all studies cited above. The most popular method, originally used by Rapp and Melville,[26] is the linear focusing in which the wave train is generated with numerous spectral harmonics that have initial phases prescribed so that all waves arrive at a certain location in phase. In realizations of this method by different authors a number of spectral shapes and widths were employed. Since nonlinear interactions between waves lead to significant spectral changes that cannot be neglected in the process of focusing of numerous harmonics, actual breaking occurs at locations different from the designed value. Nonlinear version of this method applied by Shemer *et al.*[15] enables obtaining a single steep wave at a prescribed location. In an alternative approach, steep monochromatic waves are generated by the wavemaker and undergo breaking either close to their generation location, if the initial steepness is sufficiently high, or at some randomly varying distance from the wavemaker as a result of developing instabilities.

In the present study a different approach based on the so-called Peregrine[27] breather (PB), that represents an analytic solution of the nonlinear Schrödinger (NLS) equation, was implemented. In an attempt to verify whether the PB can be obtained in a wave tank, Chabchoub *et al.*[28] observed that an initially small 'hump' in a nearly monochromatic wave train generated by a wavemaker in a tank becomes strongly amplified, so that a very steep wave is observed at some distance from the wavemaker. Shemer and Alperovich[29] demonstrated that the behavior of the wave group envelope differs significantly from that of PB. Nevertheless, the PB breather approach offers a convenient method to excite steep waves with and without breaking.

The next difference is related to the technique of velocity measurements. In all previous studies, Eulerian velocities at fixed locations beneath breaking waves were measured, mostly using Laser Doppler Velocimetry (LDV) or, more recently, Particle Image Velocimetry (PIV). We use Particle Tracking Velocimetry (PTV) that enables us to study variation in time of the horizontal coordinate of floating tracer particles and thus to obtain a varying in time

Lagrangian kinematic characteristic of the flow at the surface of a steep wave. In addition, special care is taken to accurately determine the instantaneous crest velocity.

To the best of our knowledge, only limited data on Lagrangian description the breaking process exist (see e.g. Pen et al.[30] who studied breaking of shallow water waves approaching a beach). The present experimental approach allows to study variation of the fluid particle velocity in the vicinity of the crest for nonbreaking waves, as well as prior and during the appearance of a spilling breaker, and thus to determine conditions at the inception of breaking.

## I. METHODOLOGY

The Peregrine[27] breather represents an analytic solution of the NLS equation. This equation is the simplest theoretical model describing evolution in space and time of nonlinear gravity waves propagating in deep water, valid for narrow-banded wave groups. Consider narrow banded wave group with the carrier frequency $\omega_0$ and the wave number $k_0$ that satisfy the deep-water dispersion relation $\omega_0^2 = k_0 g$. At the leading order the group can be presented as

$$\eta(x,t) = \text{Re}\left[ a \cdot e^{i(k_0 x - \omega_0 t)} \right], \tag{1}$$

where $a(x, t)$ is the complex envelope. Following Mei et al.[31] the following dimensionless scaled variables may be defined in the frame of reference moving with the group velocity $c_g$:

$$\tau = \varepsilon \omega_0 (x/c_g - t); \quad X = \varepsilon^2 k_0 x; \quad A(X,\tau) = a/a_0 \tag{2}$$

where $a_0$ is the characteristic wave amplitude, and $\varepsilon = a_0 k_0$ is the wave steepness that is the small parameter of the problem. In these variables, the evolution of the normalized envelope of the surface elevation $A(X, \tau)$ as a function of the 'slow' spatial $X$ and temporal $\tau$ variables (note that $\tau$ corresponds to negative time) can be described by the spatial version of the NLS equation:

$$-i\frac{\partial A}{\partial X} + \frac{\partial^2 A}{\partial \tau^2} + |A|^2 A = 0 \tag{3}$$

The envelope of the Peregrine breather given by

$$A(X,\tau) = -\sqrt{2}\left[1 - \frac{4(1-4iX)}{1+4\tau^2+16X^2}\right] e^{-2iX} \tag{4}$$

represents the analytical solution of (3). At the origin of the scaled coordinate system $|A(0,0)| = 3\sqrt{2}$, while $|A(X\to\pm\infty, \tau\to\pm\infty)| = \sqrt{2}$, so the maximum wave amplitude at the origin exceeds the background amplitude $\eta_0 = \sqrt{2}a_0$ by the factor of 3. The dependence of the maximum amplification factor $|A(X, 0)|/\sqrt{2}$ calculated from (4) is plotted in Fig. 1.

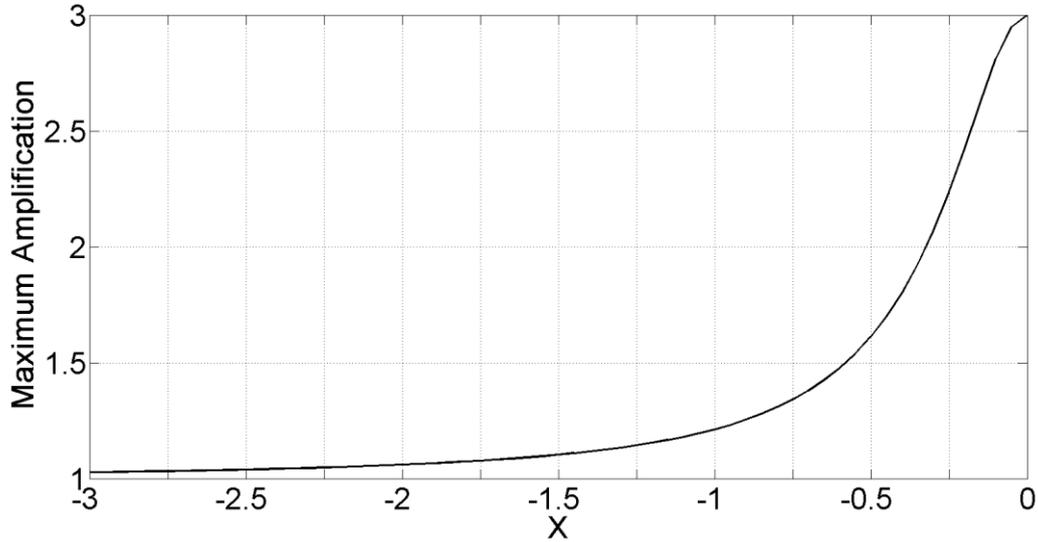

FIG. 1. Variation with the dimensionless distance $X$ of the relative envelope amplification in the Peregrine breather

Shemer and Alperovich[29] demonstrated experimentally that the spatial evolution of the PB from the initially nearly monochromatic wave train (at large negative values of $X$), to appearance of a strongly amplified wave is characterized by spectral widening. The narrow spectrum assumption of the NLS equation is thus violated resulting in quantitative behavior of the wave group envelope that significantly deviates from that of PB. They also showed that the modified NLS equation[32] in which the requirements on the spectral width are relaxed provides a better quantitative agreement between the experiment and the theory. Experiments of Shemer and Alperovich[29] indicated that acceptable quantitative agreement between the NLS solution and the experiment can only be expected as long as $X < -0.3$. As can be seen from Fig. 1, this limitation practically means the PB solution remains reasonably accurate provided that the maximum amplitude at the leading order does not exceed about $2\eta_0$. These considerations were taken into account in selection of the experimental parameters.

The experiments were performed in an 18 m long, 1.2 m wide and 0.9 m deep (water depth of 0.6 m) wave tank equipped with a programmable wavemaker. Wave energy absorbing beach is installed at the far end of the tank. Wave trains with the carrier wave period of $T_0=0.8$ s

(wave length $L_0$=1.0 m) were generated by the wavemaker according to (1), (2) and (4). Each wave train contained 70 carrier waves; tapering windows were applied at 2 periods at both ends of the train. In order to mitigate the effect of the waves reflected from the beach, measuring station was set at the distance of about $\Delta x$=8.75 m from the wavemaker. The carrier wave amplitude in all experiments was $a_0$=0.021 m ($\eta_0$=0.026 m), corresponding to $\varepsilon=a_0 k_0$=0.116. The maximum wave amplitude at the measuring location can be controlled by selection of the dimensionless coordinate of the wavemaker that in the present experiments covered the range -1.43 $\leq X_{wm} \leq$ -1.11. To obtain statistically significant data, at least 3 realizations of the wave train were excited for each one of the 9 dimensionless wavemaker coordinates within the prescribed range employed in the present experiments.

Visualization of the evolving and breaking waves was made using two identical cameras mounted on an instrument carriage. Both cameras provided synchronized video recordings at 2 Mpix resolution at the rate of 60 fps. The 1$^{st}$ camera was pointed at the wall of the tank providing records of instantaneous water contact line (Figure 1), covering an area of 64 by 36 cm$^2$ at spatial resolution of 20 pix/cm. The imaged area therefore covers about 2/3 of the carrier wave length, corresponding to the dimensionless longitudinal length of the image of 0.064. The second camera was aimed vertically down and imaged the water surface area that was twice smaller than that of the 1$^{st}$ camera in both directions (32 by 18 cm$^2$) with spatial resolution of 40 pix/cm. The centers of fields of view of both cameras were located at the dimensionless distance of $\Delta X$=0.737 from the wavemaker. It was observed that the inception of breaking occurred in the vicinity of $X$=-0.43, so that for the range of the prescribed initial conditions at the wavemaker the dimensionless location of the center of the field of view varied from $X$=-0.37 to $X$=-0.69. Figure 2 presents video clips showing simultaneously recorded variations of the water surface contact line and the movement of PTV particles, to emphasize the presentation of the video clips both frames were cropped and videos were slowed down to 30 fps. In the videos taken upstream of the breaking ($X$=-0.69, Figure 2a), as well as at the inception of the breaking ($X$=-0.43, Figure 2b), the periodic change of direction of particle's movement accompanied by the mean Stokes drift is clearly detectable. Strong horizontal acceleration of the particles can be seen in Figure 2b prior to the inception of the spilling breaker.

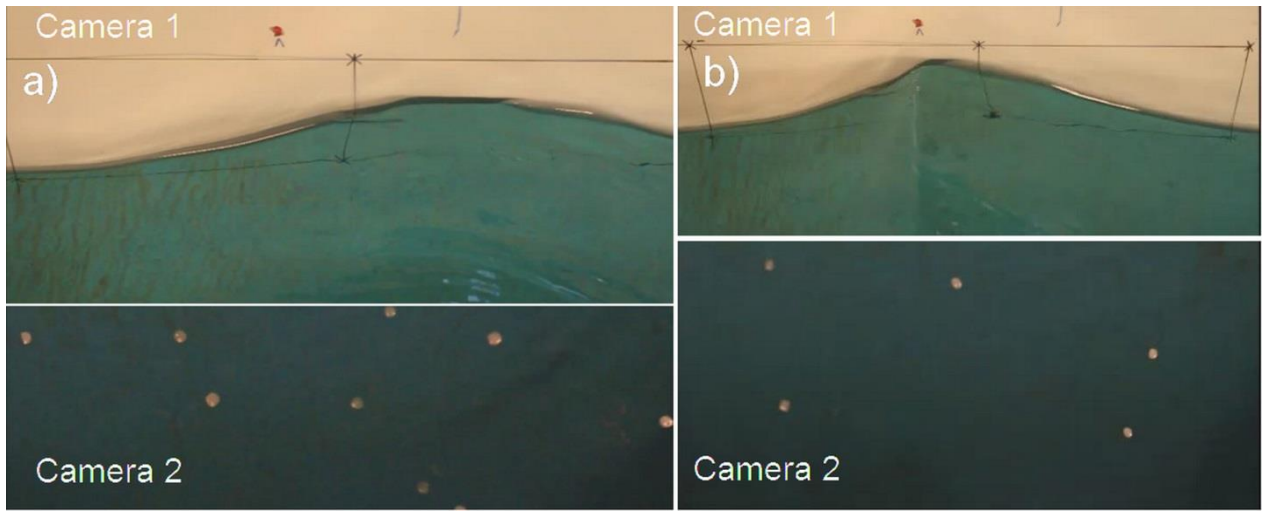

FIG. 2. Video clips (online) showing simultaneously recorded variation of the waves surface contact and the movement of the PTV particles for a) $X=0.69$ and b) $X=-0.43$ (inception of breaking). The video clips recorded at 60 fps are played at 30 fps and show only several periods around the steepest wave in the train. Frames were cropped and resized to emphasize the important flow features; actual frames' dimensions are given in the text.

Frames depicting the water contact line are presented in Figure 3 and exemplify three distinct stages of the wave train evolution. The lines visible in this Figure are drawn for image calibration purposes required because of optical distortions. Panel 3a shows the steepest wave in the train at $X = -0.65$, the wave form is nearly symmetrical and the surface is smooth. Panel 3b shows the shape of the steepest wave in the train at $X=-0.59$, the wave is characterized by a pointy shaped crest and exhibits strong front-back asymmetry. Panels 3c and 3d depict a wave during the breaking (the center of the frame is at $X=-0.43$). In panel (c) the "pointy" wave shape resembles that in panel (b), while in the panel (d) taken 0.1 s later a gentle spiller is clearly visible. The estimated crest displacement between panels (c) and (d) is 0.12 m, yielding crest velocity $c_{cr}=1.1$ m/s.

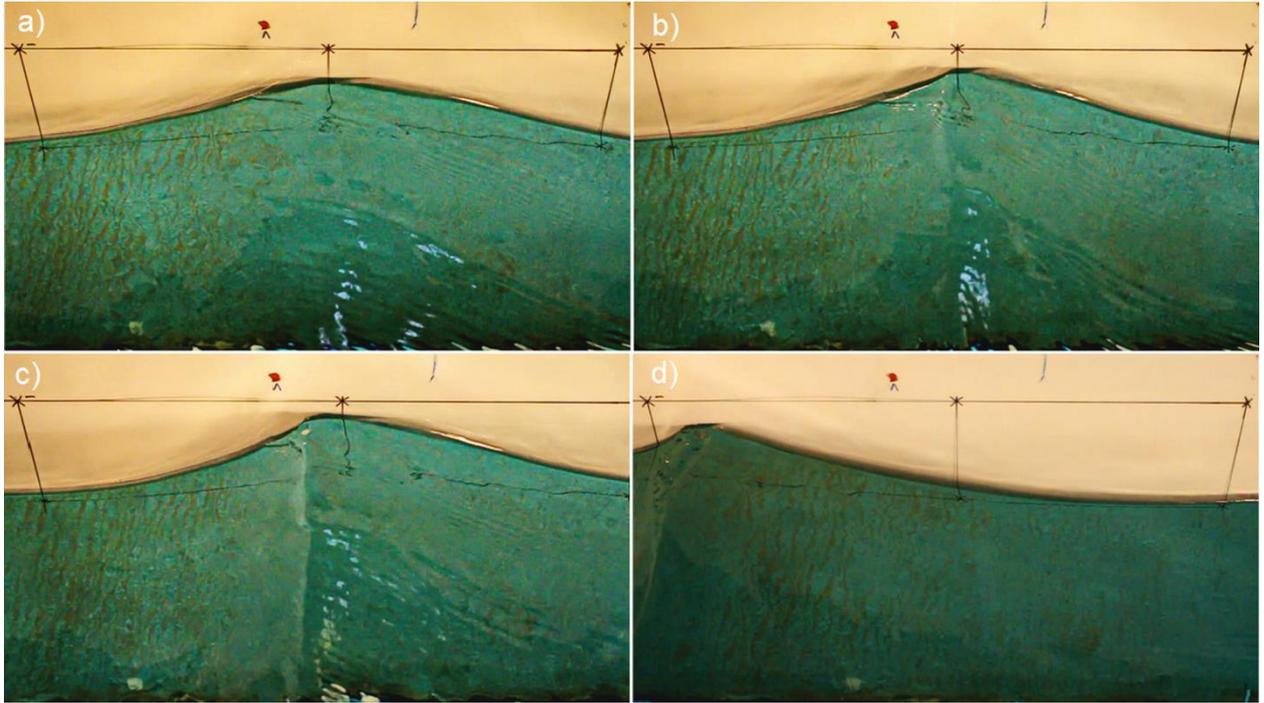

FIG. 3: a – far from breaking ($X = -0.65$), b – near breaking ($X = -0.50$), c – at the breaking ($X_{wm} = -1.38$; $X = -0.43$), d – as in c, 0.1 s later; the breaking the spiller is visible.

The water surface was seeded with buoyant particles (specific density of ~0.92) of approximately 3 mm in diameter, dispersed from above at the distance of about 1 m upstream of the imaging location. The particles then drifted towards the field of view of the cameras due to the orbital motion and mean Stokes current. Detecting positions of individual particles in each consecutive recorded video frame allowed obtaining instantaneous local horizontal velocities using Particle Tracking Velocimetry (PTV) algorithm described in details in Liberzon and Shemer.[33]

The parameters of the PTV algorithm were selected in view of the expected range of measured surface velocity variation. The characteristic scale of the water surface velocity can be estimated as $\eta_0\omega_0 = 0.204$ m/s, while the highest horizontal velocity in the vicinity of the steepest crest may be expected to exceed the linear phase velocity $c_{p0} = 1.25$ m/s. Hence the sensitivity of the PTV algorithm was adjusted to allow particle displacement up to 100 pixels, yielding the maximum detectable velocity of 1.50 m/s. It was observed that for cases where breaking occurred prior to the imaged location (i.e., for $X_{wm} = -1.11$ and $-1.13$), white scattered foam that appeared on the water surface did not allow reliable identification of the tracers. The results of the PTV measurements for these cases are therefore not presented.

To complement the video imaging, the instantaneous surface elevation fluctuations were recorded by 4 resistance type wave gages. The first gage located at $x$=8.75 m from the wavemaker and the rest are distanced from it by 10, 22.5 and 38.0 cm respectively. Data obtained by the wave gages was sampled at 1280 Hz so that each carrier wave period contained 1024 points.

## II. RESULTS

The case corresponding to $X$=-0.65 for which the breaking occurs only far downstream of the measuring location is considered first. All longitudinal velocities of individual particles at the water surface recorded during the passage of the wave train in the course of 3 runs with identical conditions are presented in Figure 4a. Each point in this Figure represents all recorded particle velocities within the images frame, showing that different runs have yielded repeatable results. The jitter in the velocity values at any given time is attributed to different instantaneous particle longitudinal positions (within $\pm 0.15 L_0$), and thus to somewhat different phases along the wave. Extreme velocities at crest and trough for a monochromatic linear wave $\pm \eta_0 \omega_0$ are plotted as well for comparison. The variability of the horizontal velocities at wave crests and troughs between consecutive waves increases notably after the appearance of the steepest wave. This qualitative difference between the behavior of the wave train before and after the steepest wave can also be observed in the corresponding temporal variation of the surface elevation during a single run plotted in Figure 4b.

The effects of nonlinearity are clearly visible in Figure 4. Crests in Figure 4b are larger than troughs, and the velocity at crests significantly exceeds $\eta_0 \omega_0$, while the values of particle velocity at the troughs are smaller than those corresponding to the linear theory. The mean Lagrangian longitudinal velocity at the water surface was found to be $\overline{U}$ =0.024 m/s, while the calculated Stokes drift velocity for a monochromatic deep water wave train is $U_{St}=\omega_0 k_0 \eta_0^2$ = 0.033 m/s. The somewhat lower measured value of the drift velocity may be attributed to the finite duration of the wave train propagating over nearly deep ($k_0 h$=3.8) water and its' essentially unsteady character, as well to experimental inaccuracy.

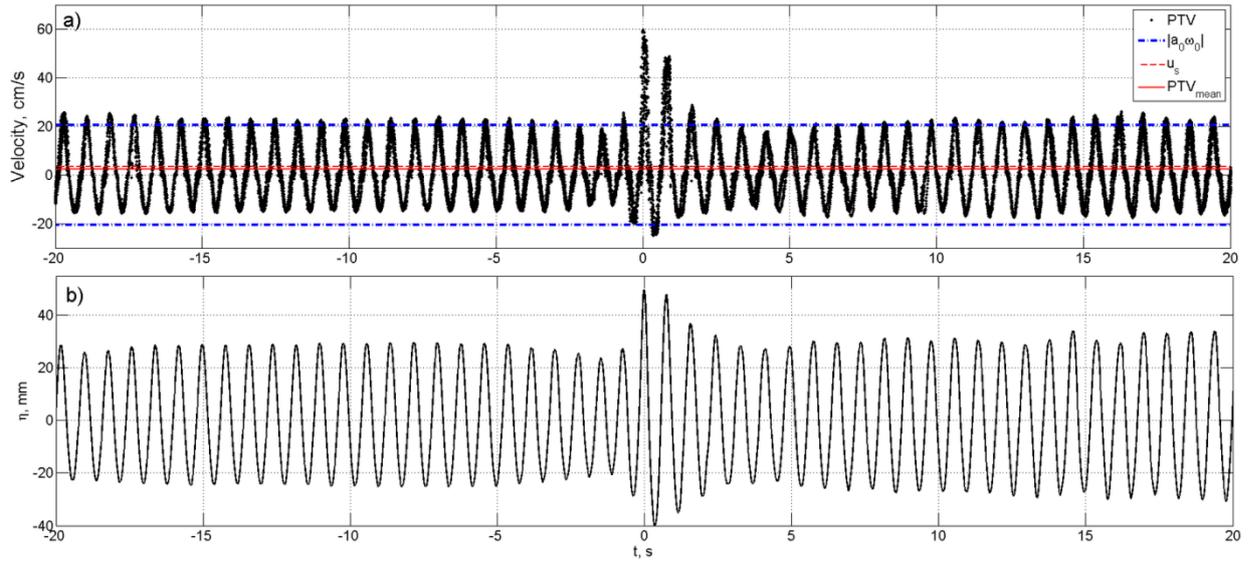

FIG. 4. Records at $X$=-0.65. a) PTV velocities at the water surface. The linear amplitude of the velocity variation, as well as the measured by PTV mean velocity and the calculated Stokes drift velocity are shown; b) Surface elevation.

Note that the maximum particle velocity at the crest of the steepest wave in Figure 4a is 0.6 m/s, well below the linear phase velocity of the carrier wave $c_{p0}$= 1.25 m/s. As already stressed, the maximum particle velocity has to be compared with the crest propagation velocity rather than with $c_{p0}$. Before proceeding to the experimental determination of crest velocities, it is instructive to take advantage of the fact that prior to breaking and at sufficiently large values of $|X|$, the wave train is expected to evolve in a reasonable agreement with the PB solution given by (2) – (4). Thus, the velocity of the steepest crest can be calculated from these expressions. The results are presented in Figure 5.

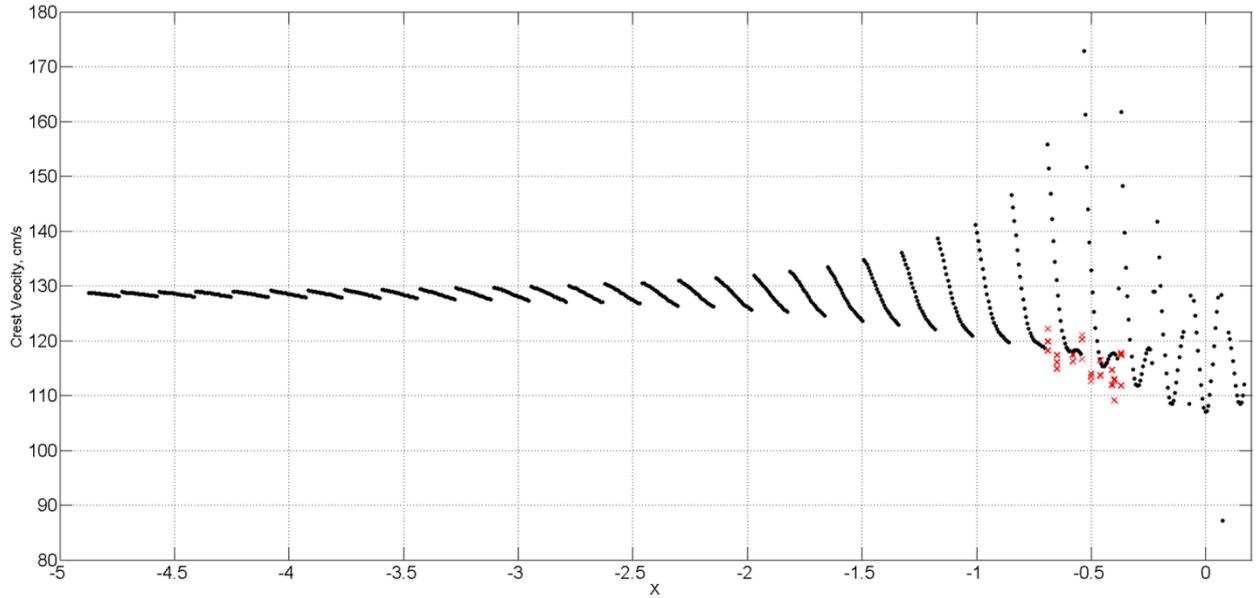

FIG. 5. Variation of the steepest crest velocity with the dimensionless longitudinal coordinate. Dots – calculations based on PB, X – experimental results.

The crest velocity of individual steep wave in the PB wave train decreases with $|X|$, with discontinuity associated with the carrier wave length and arrival of the next wave that becomes the steepest one. For $X \to -\infty$ the wave train appears as a nearly monochromatic, and the variations of the crest velocity vanish. It should be stressed, however, that the calculated PB crest propagation velocity $c_{cr}(X \to -\infty) = 1.28$ m/s, exceeding $c_{p0} = 1.25$ m/s. It can be easily shown that due to the term $e^{-2iX}$ in (4) the steepest PB crest velocity for large values of $|X|$ is given by

$$c_{cr}(X \to -\infty) = c_{p0}/(1 - 2\varepsilon^2). \qquad (5)$$

For finite carrier wave amplitude the PB wave train is therefore in fact essentially nonlinear at the leading order even at very large $|X|$ where it appears as a monochromatic wave. For values of $X$ approaching zero, the variations of the crest velocity increase, and as evident from Figure 5, the most probable crest velocities become significantly lower than $c_{p0}$, ranging from about 1.1/s to 1.2 m/s.

In the context of the present investigation, these results can only serve as estimates, and actual crest velocities have to be determined experimentally. An attempt has been made to find the actual crest velocity from the records made by the 1st video camera. Evaluation of the steepest crest velocity presented in the legend of Figure 2 indeed yields a value that is in a good agreement with the PB-based estimates. It was found however, that the very flat crest shapes as visible in the images of Fig. 2 inevitably resulted in a large scatter in the values of

$c_{cr}$ derived from the video imaging. Results reported by Qiao and Duncan[5] on crest propagation velocity that were obtained by application of a similar technique also exhibited considerable scatter. It was therefore decided to use wave gauges records to measure crest propagation velocities. For a signal of permanent shape, cross-correlation technique is used routinely for determination of the characteristic time lag for a known probes' spacing. For the conditions of the present experiment, this technique can only be used for determination of mean crest velocities in the quasi-monochromatic part of the wave train preceding the steepest wave, see Figure 4b. Two probes with the spacing of 0.255 m (about $L_0/4$) were used for this purpose. Attempts to use closer probes resulted in considerable scatter in determination of the time lag. Application of this method for numerous records with different $X_{wm}$ yields the mean estimated value of the steepest crest velocity of $c_{cr,\ exp}$=1.318 m/s. Application of the 2$^{nd}$ order Stokes correction to $c_{cr}$ calculated according to (5) results in the theoretically expected crest velocity of $c_{cr,\ th}$=1.30 m/s. The theoretical and the experimentally evaluated values of $c_{cr}$ thus agree well and exceed notably $c_{p0}$.

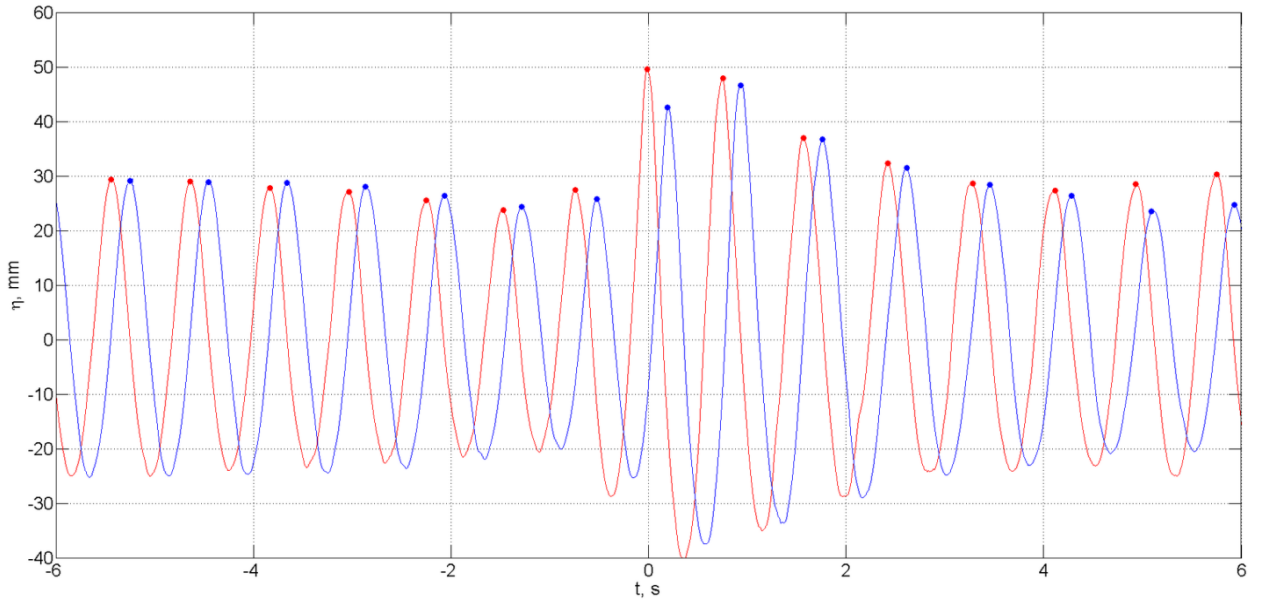

FIG. 6. Records of 2 wave gauges around the steep crest. Solid dots are the detected crests positions.

To find the propagation velocity of the steepest wave crest, the cross-correlation technique has to be replaced by direct determination of the instant of the passage of the surface elevation peak at each probe. A typical record of the temporal variation of the surface elevation in the vicinity of the steepest wave is presented in Figure 6. Second order polynomial fit was applied in the vicinity of the crest; the instant of the occurrence of the maximum surface elevation was defined according to the peak of the fit. The propagation velocities of the steepest crests at the measuring location determined in this way for all experimental runs are also plotted in Figure 5. There is a reasonable agreement between the measured crest velocities and estimates based on the PB.

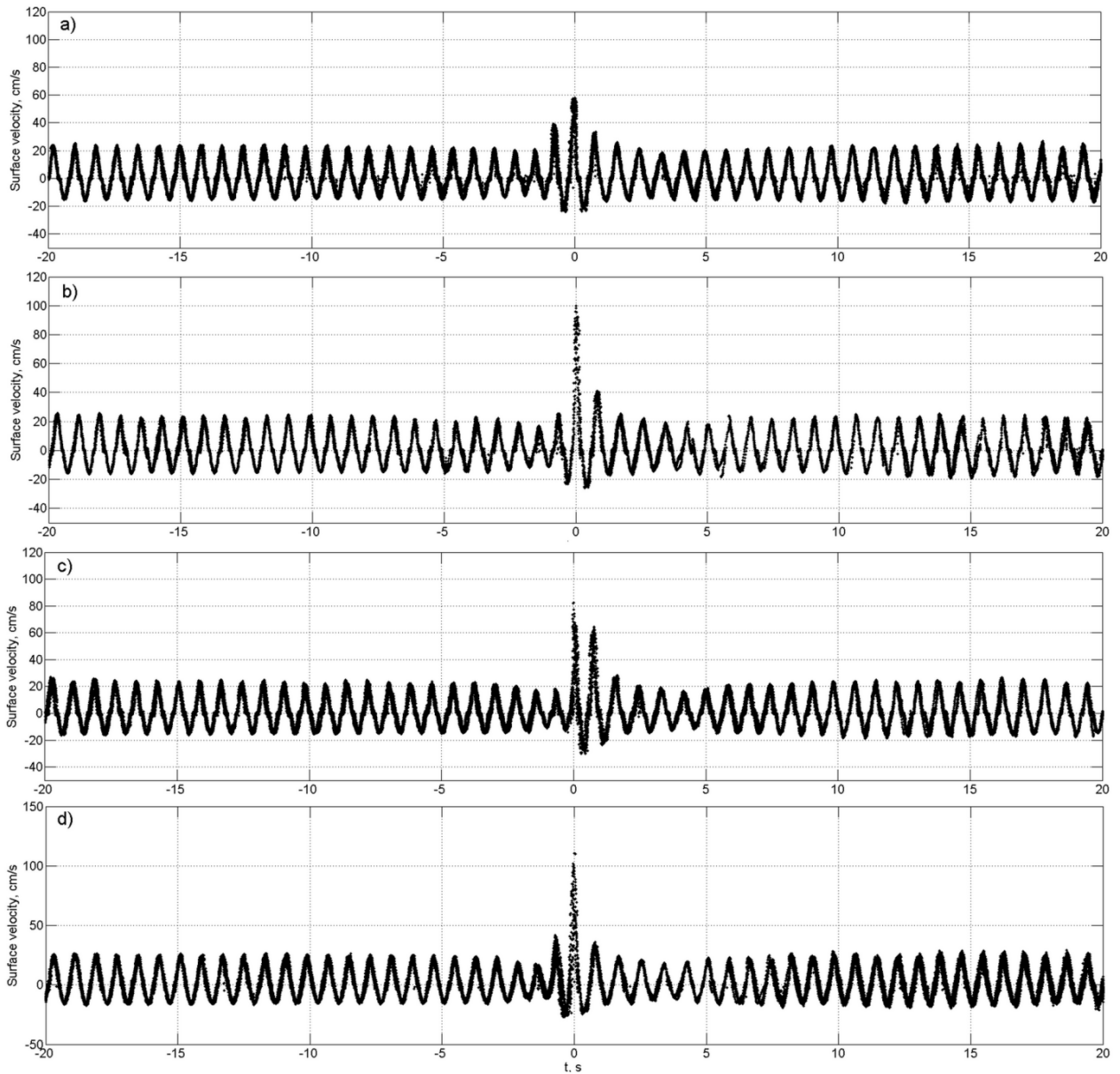

FIG. 7. The Lagrangian velocities at the water surface at a) $X=-0.69$, b) $X=-0.50$, c) $X=-0.46$, d) $X=-0.43$.

The measured during the passage of the wave train Lagrangian horizontal velocities are presented in Figure 7 for 4 dimensionless coordinates of the imaging window. The corresponding wave gauge records of the surface elevation variation are given in Figure 8.

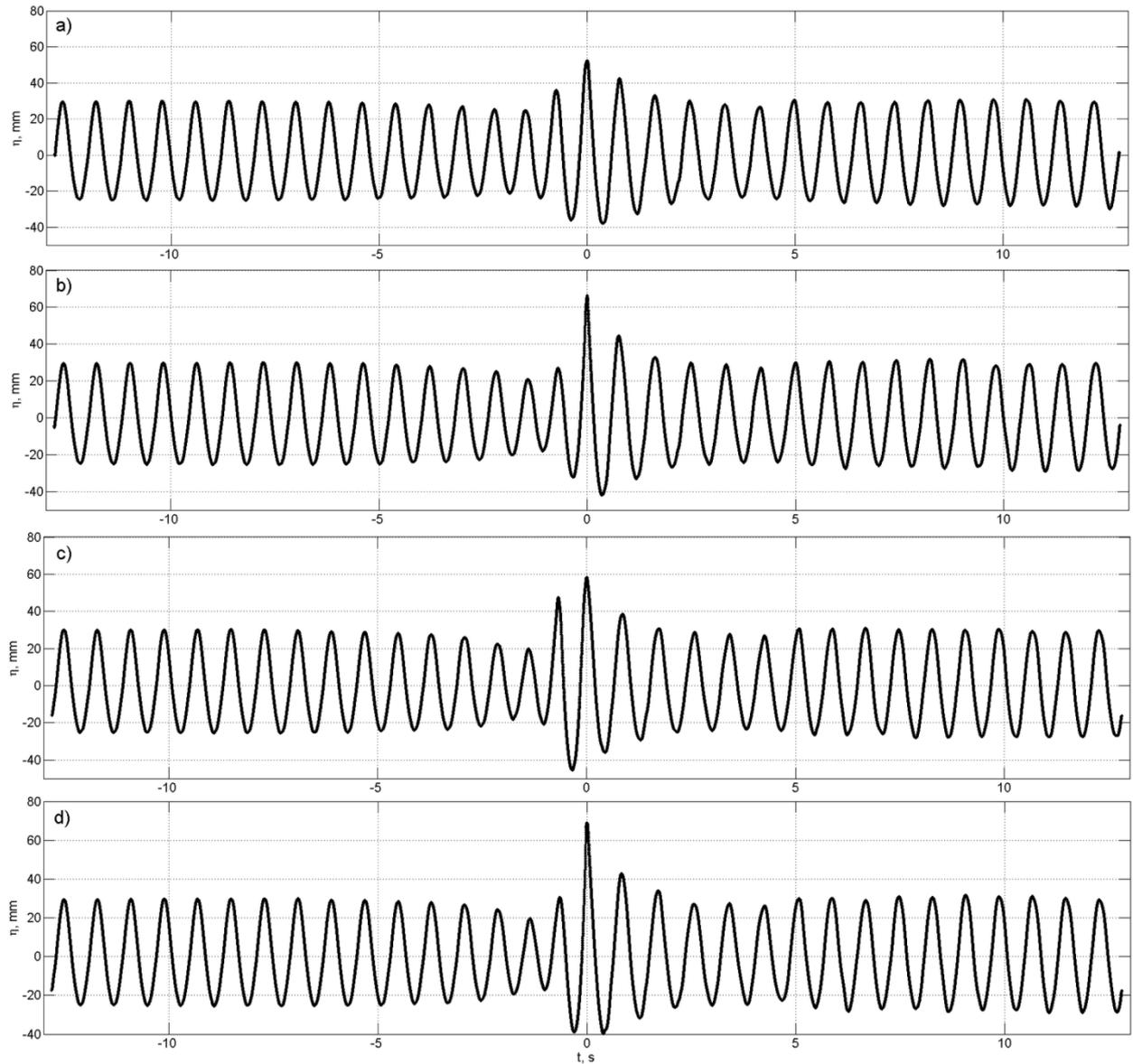

FIG. 8. Surface elevation measured by a wave gauge in a single run. Coordinates as in Figure 7.

The increase in both the maximum crest velocity and the crest elevation with approaching the inception of breaking at $X=-0.43$ is obvious. The increase is, however, non-monotonic, in qualitative agreement with Figure 5. Both the maximum horizontal velocity at the crest and the crest height in Figures 7c and 8c are lower than in Figures 7b and 8b, respectively. This is due to the fact that the local maximum values depend also on the relative phases of the slowly varying complex group envelope and of the carrier wave. For $X=-0.46$, these phase relations result in appearance of 2 steep peaks with comparable heights that are lower than those observed at locations where the extreme values correspond to the envelope phase close to zero.

Comparison of Figures 7 and 8 reveals that the relative amplification of the horizontal velocity at the steepest wave crest as compared to that at the background carrier wave is significantly stronger that the corresponding crest heights ratio. The stronger amplification of the horizontal velocity of water particles at steep crests can be attributed to the contribution of the 2$^{nd}$ and higher order bound waves. For surface elevation, this contribution manifests itself in crests larger than troughs. As discussed recently in [14], higher order terms associated with the bound waves contribute even more to the horizontal velocity at the crest of a steep wave due to the enhanced weight of the higher frequency harmonics.

A closer look at the velocities in the vicinity of the steepest wave is taken in Figure 9. The measured crest velocities are also plotted in this Figure. Away of breaking (Figure 9a), the steepest crest velocity considerably exceeds the maximum recorded surface horizontal velocity at the surface. Closer to the breaking location (Figure 9b) maximum water particle velocities approach that of the crest. At the inception of breaking (Figure 9c), the maximum horizontal fluid particle velocities at the surface and the crest velocities become virtually identical. Note the enhanced scatter around the steep crests. This scatter is attributed to the strong acceleration that particles at the surface undergo when approaching crest and trough of the wave. The differences in the particle velocities at the same instant but somewhat different longitudinal position within the imaged frame become therefore more pronounced.

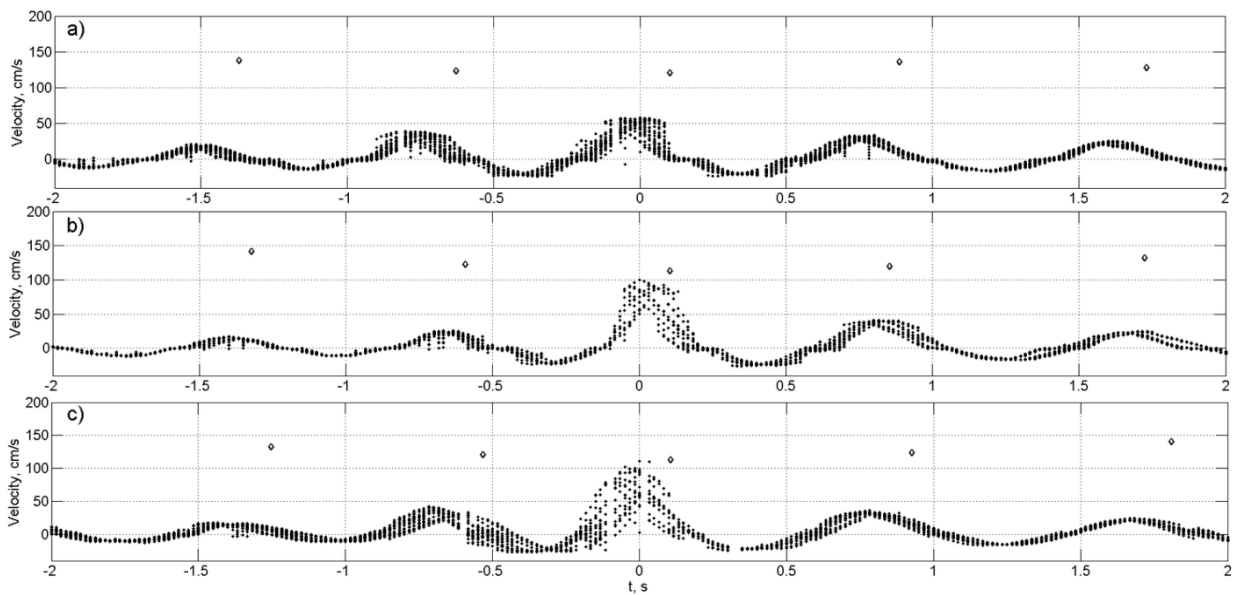

FIG. 9. Comparison of Lagrangian particle velocities in the vicinity of the steepest peak and the measured crest velocities: a) *X*=-0.69; b) *X*=-0.50; c) *X*=-0.43.

The PTV technique employed in the present study enables estimates of the Lagrangian horizontal accelerations. Strong acceleration of the tracing particles associated with the steepest crest is apparent in the video clips presented in Figure 2. Ensembles of instantaneous horizontal accelerations $a_{h,L}$ for all particles for three experimental conditions are presented in Figure 10. The linear estimate of the range of variation of the horizontal acceleration given by $\pm\eta_0\omega_0^2=\pm160.4$ $cm/s^2$ is denoted by a broken line. For the quasi-monochromatic part of the wave train, the range of variation of the Lagrangian horizontal acceleration only slightly exceeds the linear estimate. As expected the phase of $a_{h,L}$ leads that of the surface elevation and the horizontal velocity by $\pi/2$. Note also that for that part of the train, the measured dependence of $a_{h,L}(t)$ is symmetric with respect to the horizontal axis, with the mean value close to zero. For the steepest waves in the train, however, the nonlinear contributions to the Lagrangian horizontal acceleration by both high frequency bound waves and convective acceleration terms becomes significant, so that the maximum absolute values of $a_{h,L}$ notably exceed the linear estimate. The vertical symmetry is retained as long as the amplification is not too strong, see Figure 10a. Maximum acceleration increases as the steepest crest height grows, so at the inception of breaking (Figure 9c) the maximum Lagrangian acceleration that is observed prior to the highest crest exceeds the linear estimate by a factor of about 5.

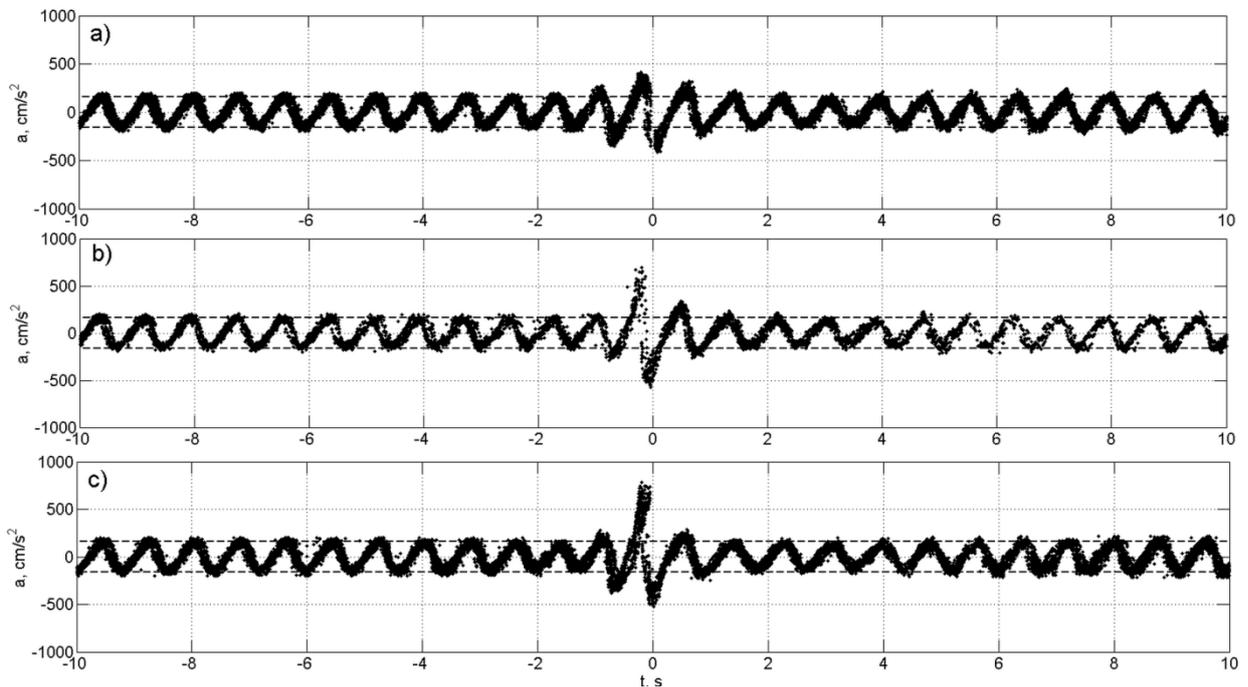

FIG. 10. Temporal variation of the Largangian horizontal acceleration: a) $X$=-0.69; b) $X$=-0.50; c) $X$=-0.43. Dashed lines denote $\pm\eta_0\omega_0^2$=+/-160.4 $cm/s^2$

It is interesting to note that while for linear deep-water waves the amplitudes of horizontal and vertical accelerations are identical, it is not so for the strongly nonlinear steep waves. As mentioned above, all available theoretical and experimental results indicate that the vertical

Lagrangian acceleration at the crest of very steep wave does not exceed about $g/3$. The reported maximum horizontal accelerations of fluid particles at the surface of very steep waves exceed this value for all cases examined.

**III. CONCLUSIONS**

Contrary to earlier investigations of breaking waves, steep waves in the present study were excited using an essentially nonlinear mechanism associated with the solution of the nonlinear Schrödinger (NLS) equation known as Peregrine Breather.[27] This mechanism causes an initially small 'hump' in the envelope of a nearly monochromatic wave train to grow along the tank, eventually attaining crest height that at the leading order exceeds the amplitude of the background carrier wave by a factor of 3. It was experimentally demonstrated [29] that this amplification is in fact never attained for water gravity waves due to spectral widening that renders the NLS equation invalid. Nevertheless, it can be concluded from these results that as long as the steepest crest height does not exceed about twice the carrier wave amplitude, the Peregrine analytical solution remains reasonably accurate. It is demonstrated here that the PB approach to studying breaking waves has two important advantages as compared to linear wave focusing that was routinely applied in the past. First, within the limits of validity, the availability of theoretical solution allows to obtain fairly accurate estimates of the expected wave parameters, thus facilitating the design of the experiment. Second, for a given carrier wave, the steepest wave crest height at the measuring location can be accurately controlled by varying a single governing parameter – the dimensionless location of the wavemaker in the frame of coordinates appropriate for PB. This approach thus made possible to investigate kinematics of a steep wave with gradually varying crest height, up to the inception of a spilling breaker.

Kinematics of steep waves on the verge of breaking and beyond was studied by two synchronized video cameras and a set of wave gauges distributed along the tank. The 1[st], side-looking camera, provided records of the temporal variation of the contact line at the tank wall, thus enabling identification of the inception of breaking. The 2[nd] camera imaged the water surface from above and provided records of the instantaneous location of tracing particles floating at the water surface. Application of the PTV algorithm provided data on the Stokes drift and the instantaneous Lagrangian horizontal velocity, as well as on acceleration of the tracers.

The goal of this study was to determine the criterion for the inception of breaking. Two criteria based on physical considerations were considered: dynamic and kinematic. Since Phillips[12] dynamic criterion by apparently cannot be satisfied for Stokes waves[13] as well as for a broad-banded wave group,[14] special attention was given to the kinematic criterion.. It is argued here that this criterion relates the maximum velocity of a fluid particle at the crest of the steep wave to the crest velocity. It was demonstrated that the determination of the actual crest velocity and its relation to the phase velocity of the carrier wave $c_{p0}$ requires particular attention.

At all stages of its evolution, PB consists of a localized 'hump' that has duration of few carrier wave periods; the rest of the wave train appears as a nearly monochromatic wave. It was demonstrated, however, that the essentially nonlinear character of PB given by (4) results in the effective phase velocity of this quasi-monochromatic part somewhat exceeding that of a Stokes wave. This observation was confirmed by measurements. The situation is different in the vicinity of the 'hump'. Simple analysis based on the analytical expression for the PB demonstrated that the propagation velocity of a steep crest is considerably reduced as compared to the phase speed of the carrier wave $c_{p0}$. The actual crest velocity is strongly affected by the finite width of the wave train spectrum, with the instantaneous location of the crest resulting from constructive interference of numerous spectral components contributions with different phases. The slower than $c_{p0}$ velocity of propagation of the steepest crest is therefore mostly determined by linear effects and the free wave part of the spectrum.. At the same time both the horizontal and the vertical velocity components at water surface in the vicinity of the steep crest are strongly affected by the higher order contributions.[14]

These theoretical estimates were supported by experiments. Two independent methods to measure crest velocity were applied in this study. The first method was based on estimating the rate of crest displacement from the video images acquired by the 1[st] camera. While this method yielded reasonable estimates, the scatter of the results that stemmed from the flat shape of the crest as well as from optical distortions was quite substantial. It was decided therefore to determine the steepest crest propagation velocity from wave gauges measurements that were acquired at 1280 Hz/channel. Even at this relatively high sampling rate, the distance between the probes of about ¼ of the carrier wave length was required to obtain accurate enough results, at the expense of ability to catch the short-scale variations in the crest velocity.

The measured steep wave propagation velocities in the present experiments were in good agreement with calculations based on the PB. As long as no breaking was observed, the propagation velocities of the steepest wave crest exceeded the maximum recorded surface particle horizontal velocity. An essentially nonlinear growth of the horizontal velocities of water particles at the surface with the increase in crest height was observed. At the inception of a spilling breaker, the horizontal velocities at the surface attain that of the crest. This observation confirms the validity of the kinematic criterion for inception of breaking.

To attain high velocities at the crest, the Lagrangian accelerations of water particles grow significantly reaching very high values. It is therefore plausible to assume that once the material horizontal velocity at the surface attains that of the crest (at the inception of breaking) and then exceeds it as the wave breaking process evolves to advanced stages, the accumulation of mass at the crest leads to formation of a 'bulge' on the forward face of the crest of a gentle spilling breaker as reported by Dunkan *et al.*[4]

**ACKNOWLEDGEMENTS**

This study was supported by a grant # 2010219 from US-Israel Binational Science Foundation.